\documentclass[
 reprint,
 amsmath,amssymb,
 aps,
]{revtex4-2}

\usepackage{graphicx}
\usepackage{dcolumn}
\usepackage{bm}

\begin{document}

\preprint{APS/123-QED}

\title{Precise pulse shaping for quantum control of strong optical transitions}
\author{Yudi Ma$^1$}
\author{Xing Huang$^1$}
\author{Xiaoqing Wang$^2$}
\author{Lingjing Ji$^1$}
\author{Yizun He$^1$}
\author{Liyang Qiu$^1$}
\author{Jian Zhao$^1$}
\author{Yuzhuo Wang$^1$}
\author{Saijun Wu$^1$}%
\email{saijunwu@fudan.edu.cn}
\affiliation{%
$^1$Department of Physics, State Key Laboratory of Surface Physics and Key Laboratory of Micro and Nano Photonic Structures (Ministry of Education), Fudan University, Shanghai 200433, China.\\
$^2$State Key Laboratory of Quantum Optics and Quantum Optics Devices, Institute of Laser Spectroscopy, Shanxi University, Taiyuan 030006, China.
}%


\date{\today}

\begin{abstract}
Advances of quantum control technology have led to nearly perfect single-qubit control of nuclear spins and atomic hyperfine ground states. In contrast, quantum control of strong optical transitions, even for free atoms, are far from being perfect.  Developments of such quantum control appears to be bottlenecked by available laser technology for generating isolated, sub-nanosecond optical waveforms with sub-THz programming bandwidth. Here we propose a simple and robust method for the desired pulse shaping, based on precisely stacking multiple delayed picosecond pulses. Our proof-of-principal demonstration leads to arbitrarily shapeable optical waveforms with 30~GHz bandwidth and $100~$ps duration. We confirm the stability of the waveforms by interfacing the pulses with laser-cooled atoms, resulting in  ``super-resolved'' spectroscopic signals. This pulse shaping method may open exciting perspectives in quantum optics, and for fast laser cooling and atom interferometry with mode-locked lasers.
\end{abstract}
\maketitle




\section{Motivation}

\begin{figure*}[ht]
\centering
\includegraphics[width=\textwidth]{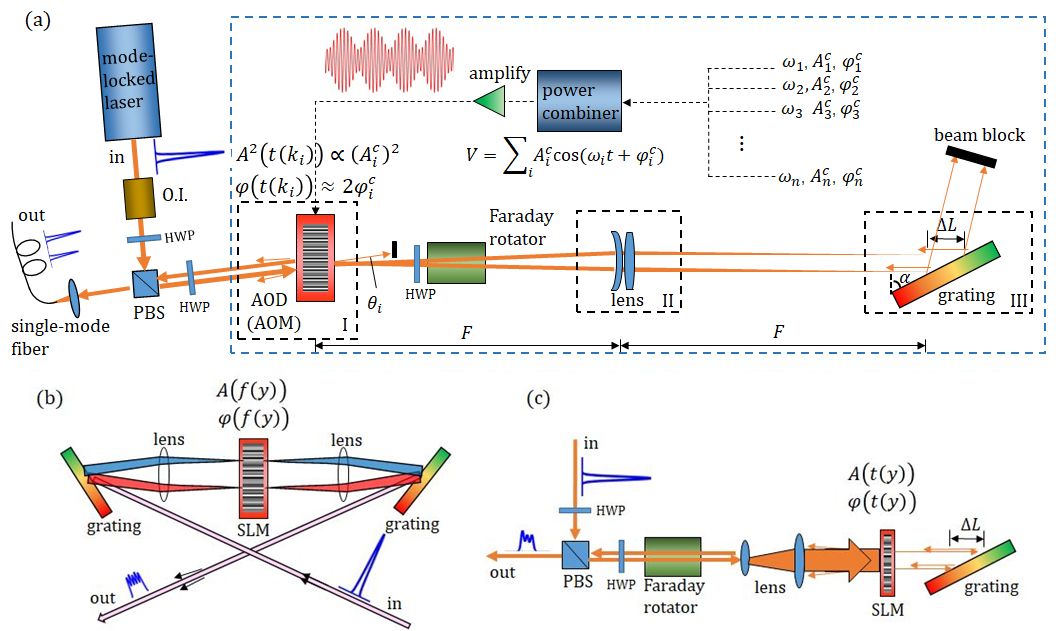}
\caption{(a) Schematic for the diffractive multi-delay generation based picosecond pulse shaping setup in this work. (b) Schematic for a representative Fourier transform pulse shaping setup. (c) Schematic for a representative direct space-to-time pulse shaping setup. PBS: polarization beam-splitter. AOD: Acoustic Optical Deflector. AOM: Acoustic Optical Modulator. O.I.: Optical Isolator.  HWP: half-wave plate. SLM: Spatial light modulator.}
\label{fig:setup}
\end{figure*}


Quantum control on short lived optical transitions has historically been limited by the lack of suitable programmable light sources.  In this paper we extend and adapt techniques from ultrafast science~\cite{shaperTutorial2010,Weiner2011Review} to be compatible with controlling isolated electric dipole transitions in free atoms. Specifically we realize shaped 100~ps, 795~nm pulses with 30 GHz bandwidth which are used to coherently drive loss from optically trapped $^{87}$Rb atoms on the D1 line.  This first demonstration opens the door to more sophisticated quantum control applications such as implementing precise phase gates for cooperative emission control in atomic gases~\cite{scully2015,YizunHe2019}.

To clarify the technical challenges associated with the quantum controls that motivate this work, we generally consider an optical ``spin'' defined on a strong transition of a free atom between its ground state $|g\rangle$ and an excited state $|e\rangle$. It is well-known that the dynamics of such optical spin can be represented by a state vector on a Bloch sphere~\cite{ScullyBook}. The quantum state can thus in principle be precisely controlled through electric dipole interaction with the electric field of an optical pulse, with the complex field ${\bf E}(t)$ at the positive frequency and the associated Rabi frequency $\Omega(t)={\bf E}(t)\cdot {\bf d}_{eg}e^{i\omega_{eg}t}/\hbar$.  Here $\omega_{e g}$ and ${\bf d}_{e g}$ are the transition frequency and the dipole matrix element respectively. 
However, precise control of the optical spins is substantially more challenging than the microwave control of nuclear spins~\cite{NMRAdiabatic2001}. To start with, a prerequisite for high precision 2-level control is to maintain a ``clean'' system free of uncontrolled multi-level couplings. To avoid the couplings or even photo-ionization in real atoms, moderate peak field $|{\bf E}_{\rm peak}|$ and peak Rabi frequency $\Omega_{\rm peak}$ are preferred. Meanwhile, simple but meaningful controls require a long enough duration $\tau_{\rm c}$ at $\pi/\Omega_{\rm peak}$ level, such as for coherent population inversion. The combined requirements for real atoms suggest that to achieve precise control of an optical transition, control over long duration with narrow-band or picosecond pulses are preferred over ultrafast pulses. But techniques for shaping of picosecond pulses are relatively underdeveloped~\cite{shaperTutorial2010,Weiner2011Review}. On the other hand, for a strong transition with a typical $\Gamma_e\sim 2\pi \times 10$~MHz linewidth, optical pulses with $\tau_{\rm c}\ll 1/\Gamma_e$ are challenging to produce by modulating continuous wave (CW) lasers, while control with multiple pulses from mode-locked lasers or frequency combs~\cite{HanschCombSpecReview2019} would require a pulse train with repetition $T_{\rm rep}\ll 1/\Gamma_e$, also beyond the standard technology. Finally, for a laser beam with intensity $I\propto |{\bf E}|^2$ and $\Omega({\bf r},t) \propto \sqrt{I({\bf r},t)}$, intensity inhomogeneity across the sample and shot to shot variation over time translate directly into the spin control error, leading to degraded average control fidelity if a highly uniform laser intensity profile with temporary power stability cannot be maintained.

The pulse shaping method to be introduced in this work is motivated by the possibility of extending the composite pulse techniques as applied in Nuclear Magnetic Resonance (NMR)
~\cite{NMRAdiabatic2001,Navin2005grape, Odedra2012,geometricAspect2012,GenovFieldError2014,NatCommDu2015,Chuang2016} to the optical domain, so as to achieve error-resilient precise control of strong optical transitions in a stable and robust manner. It is well-known in the quantum control community~\cite{geometricAspect2012,GenovFieldError2014,NatCommDu2015,Chuang2016} that intensity errors during the 2-level control can be suppressed by shaping the control pulses with optimized amplitude and phase functions. We shall characterize the shaped waveform with a modulation bandwidth $\delta f_{\rm M}$ and duration $\tau_{\rm c}$. Since flexible shaping requires a large $\delta f_{\rm M} \times \tau_{\rm c}$ product, useful shaping for sub-nanosecond $\tau_{\rm c}$ controls requires arbitrarily programmable bandwidth $\delta f_{\rm M}$ beyond 10~GHz and into the sub-THz regime. This modulation bandwidth is not well suited to CW laser modulation technology~\cite{accMetcalf2007,Gould2016,YizunHe2019}. On the other hand, as will be clarified further, the well-developed ultrafast pulse shaping techniques~\cite{shaperTutorial2010,Weiner2011Review} are typically not suitable for the relatively narrow-band control. In this work we propose a precise pulse shaping method, with transform-limited picosecond pulses as inputs, that supports the generation of arbitrarily programmable waveforms with duration $\tau_{\rm c}$ approaching nanoseconds and with sub-THz modulation bandwidth $\delta f_{\rm M}$ only limited 
by the bandwidth $\delta f_{\rm in}^{\rm L}$ of the input picosecond pulse. The proposed setup, in its basic form, is schematically shown in Fig.~\ref{fig:setup}(a). The basic idea is to perform precise pulse shaping in the time domain~\cite{OriginOfDSTPS1992,WeinerDST1999, diffractiveAWGOptLett2010, CompactDST2011}, by coherently stacking multiple sub-pulses with arbitrarily programmable amplitudes, relative phases, and delays. To generate the multiple delays, acoustic-optical deflections in the ``double-pass'' configuration are implemented to create stable, co-propagating multiple optical delay lines which are interferometrically coupled to a single mode output. Comparing with previous time-domain shaping methods, our method divides the initial pulse in ${\bf k}$ space instead of e.g. cutting wavefronts in real space, leading to stable single-mode output coupling efficiency which is insensitive to the shaping parameters. Similar to Acoustic Optical Programmable Dispersive Filter (AOPDF)~\cite{AOTFWeinerOL1993, AODispersiveFilter1997}, the multi-delay based pulse shaper is a linear optical filter, but supports orders of magnitude longer delays for control with long pulse duration $\tau_{\rm c}$. Specifically, the delays supported by ``retro-diffraction'' from a large-area grating can be a few hundred of picoseconds, while delays beyond nanoseconds can be implemented with additional gratings combined with proper beam waist managements. 
The passive phase stability for the co-propagating delay lines is protected by common-mode rejection of vibration noise. Long term waveform stability can be interferometrically monitored by a CW laser injected into the same optical paths. 

In the following we first outline the operation principle of the new pulse shaper scheme. We then discuss a proof-of-principle demonstration of the method by achieving  $\delta f_{\rm M}\approx 30$~GHz arbitrarily shapeable pulses with control duration $\tau_{\rm c}$ up to 100~ps in a ``small signal'' regime. We realize a first demonstration of coherent interaction between shaped picosecond laser pulses with laser-cooled atoms. Owning to the excellent phase stability of the shaper, we demonstrate remarkably sharp, ``super-resolved'' spectroscopic features with frequency resolution beyond the transform limit of the pulses. In the discussion section we quantify the performance limitations to the shaper scheme, in particular, we outline a path toward operating the programmable shaper at the power efficiency limit with long-term waveform stability. We then discuss the relation between this scheme with traditional frequency and time-domain shaper schemes, and clarify the unique potential of our scheme for generating isolated, wide-band shapeable waveforms for high-precision quantum control of strong optical transitions.

\section{Programmable pulse shaping by diffractive multi-delay generation}\label{Sec:shaper}
We consider pulsed picosecond optical input at a central laser frequency $\omega_{\rm L}$ with a spectral bandwidth $\delta f_{\rm in}^{\rm L}$ and temporal duration $\tau$, with the positive frequency part of the electric field expressed as ${\bf E}_{\rm in}(t)$. Spatially the input is in a Gaussian mode with collimated waist $w$ and wave-vector ${\bf k}_{\rm in}$. As in Fig.~\ref{fig:setup}(a), the input pulse is deflected (Bragg-diffracted) by an acoustic-optical deflector (AOD) (I) driven by multiple frequency rf source with angular frequency $\omega_i$ ($i=1,...,N$).   For each $\omega_i$ sound wave, the acoustic-optical deflection leads to sub-pulse ${\bf E}^{(1)}_i= A_i e^{i\varphi_i^{\rm c}} {\bf E}_{\rm in}$ at deflection angle $\theta_i=\omega_i/|{\bf k}_{\rm in}| v_{\rm s}$ ($v_{\rm s}$ is the AOD speed of sound), with diffraction amplitude $A_i\propto A_i^{\rm c}$ and phase $\varphi_i^{\rm c}$ controlled by the driving rf signal at the $\omega_i$ frequency, in the ``small signal'' and linear regime. 

After the AOD, a sub-pulse ${\bf E}^{(1)}_i$ is focused by an aberration-corrected wide-field lens (II) at a distance $F$ away into its $w_{\rm G}= \lambda F/\pi w$ Gaussian waist. Here $F$ is the effective focal length and $\lambda=2\pi/|{\bf k}_{\rm in}|$ is the central wavelength of the laser. To generate tunable optical delays, a large area grating (III) is centered to the focal plane to ``retro-diffract'' the pulse back, in a ``broadband regime'' with grating diffraction bandwidth $\delta f_{\rm G}\gg \delta f_{\rm in}^{\rm L}$ (see Sec.~\ref{SecMDMLimit}), to its approximately time-reversed wavefront. The $\omega_i$-independent diffraction is helped by centering the AOD to the input focus of the lens to obtain $\omega_i$-independent incident angle $\alpha={\rm asin}(\lambda/2 d)$ ($d$ is the grating constant) toward the grating. To achieve efficient retro-diffraction in this configuration~\cite{LittrowScholten2001}, the polarization of the input beam needs to be optimized, and high density gratings with $d<\lambda$ are preferred to ensures single order diffraction. 
By introducing the $\omega_i$-dependent extra optical path $\Delta L_i=\theta_i {\rm tan}(\alpha) F$, the retro-diffraction induced optical delay $\tau_i=2 \Delta L_i /c$ is expressed as
\begin{equation}
    \tau_i=\frac{\omega_i \lambda}{v_{\rm s}} \frac{F}{\pi c \sqrt{4d^2/\lambda^2-1}},\label{equ:delay}
\end{equation}
without introducing delay-dependent phase shifts.

Aided by the time-reversal symmetry, the retro-diffracted ${\bf E}_i^{(1)}$ back through the $F$-lens is deflected again by the $\omega_i$ sound wave in AOD, with nearly identical deflection efficiency as that for the first deflection. The complex electric field after the 2nd AOD can be expressed as ${\bf E}_{i,\rm out}=A_i e^{i (\varphi_i^{\rm c}+\delta \varphi_i^{\rm c})} {\bf E}_i^{(1)}$. Here $\delta \varphi_i^{\rm c}=\delta \varphi_i^{\rm c}(\omega_i)$ accounts for sound wave phase change during the $\omega_i$-dependent optical delay. The wave-vector is shifted back to the $\omega_i$-independent ${\bf k}_{\rm out}=-{\bf k}_{\rm in}$ for the single-mode output coupling. Taking into account an overall loss coefficient $\kappa$ including those due to the grating and fiber couplings, the shaped composite pulse can be expressed as
\begin{equation}
    \begin{array}{l}
{\bf E}_{\rm out}(t)=\kappa \sum_i^N 
    {\bf E}_{i,\rm out}(t),\\
    ~~~~~~~~~~~~~=\kappa \sum_i^N A_i^2 e^{i \varphi_i } {\bf E}_{\rm in}(t+\tau_i).
\end{array}\label{equ:inOut}
\end{equation}

To arrive at Eq.~(\ref{equ:inOut}) we have ignored the rf frequency shifts to the picosecond pulses~\cite{notesFshift}
. With $A_i\propto A_i^{\rm c}$ and $\varphi_i=2\varphi_i^{\rm c}+\delta \varphi_i^{\rm c}$ in the linear regime of the shaper operation, it is straightforward to generate arbitrary $\{A_i,\varphi_i,\tau_i\}$ sub-pulse arrays with the $\{A_i^{\rm c},\varphi_i^{\rm c},\omega_i\}$ rf control signals as prescribed by Eqs.~(\ref{equ:delay})(\ref{equ:inOut}). For the input pulse with duration $\tau$, the overall duration $\tau_{\rm c}$ of the shaped waveform is limited to $(\tau_{\rm d})_{\max}+\tau$, with the maximum delay $(\tau_{\rm d})_{\max}$ specifying the range of $\{\tau_i\}$ supported by the grating retro-diffraction (Sec.~\ref{SecEffLimit}). As for all the linear optical filters, the filter efficiency for the single-mode $N$ sub-pulses with nearly equal amplitudes and arbitrarily programmable phases is fundamentally limited to $\eta_{\rm E}\sim |\kappa|^2 \sum |A_i|^4 \sim 1/N$ in terms of energy, and $\eta_{\rm P}= |\kappa|^2 \langle |A_i|^4 \rangle \sim 1/N^2$ in terms of peak power. With ${\bf E}_{\rm in}(\omega)$ in the frequency domain, it is also convenient to rewrite Eq.~(\ref{equ:inOut}) as ${\bf E}_{\rm out}(\omega)= s(\omega) {\bf E}_{\rm in}(\omega)$, with the filter function
\begin{equation}
    s(\omega)=\kappa \sum_i^N A^2_i e^{i\varphi_i}e^{-i\omega\tau_i}.\label{equ:inOutf}
\end{equation}


\section{Experimental demonstration}\label{Sec:Exp}
We demonstrate the Fig.~\ref{fig:setup}(a) scheme by shaping picosecond pulses from a mode-locked laser (Spectro-Physics Tsunami system) with $\lambda=795$~nm, transform-limited $\tau\approx11$~ps pulses at $f_0=80$~MHz repetition rate. The beam waist is adjusted to $w=1.5~$mm for the acoustic-optical deflection, which is realized by an acoustic-optical modulator~(AOM) with 80~MHz central frequency and 20~MHz bandwidth. To repeat the pulse shaping at MHz rate, the AOM is driven by multi-frequency rf at $\omega_i=2\pi\times (f_0+n_i \Delta f )$ with $n_i$ and $\Delta f$/MHz set as integers. The $\omega_i$ signals from synthesized rf sources (NOVATech 409B), phase-locked to the $f_0=80~$MHz reference, are combined with a multi-port splitter (Minicircuits 15542 ZFSC-24-1). In typical experiments with fixed $\omega_i$ combinations, $A_i^{\rm c}$ and $\varphi_i^{\rm c}$ are rapidly programmed into the rf sources, amplified to update the AOM sound waves in less than a microsecond. The ${\bf E}_{\rm out}$ pulses at $\varphi_i^{\rm c}$-specific times are post-selected by an AOM-based pulse picker (not shown in Fig.~\ref{fig:setup}(a)), with $\sim 20\%$ pulse picking efficiency and at a lower $T_{\rm rep}=1/\Delta f$ repetition.  With speed of sound $v_{\rm s}=4260$~m/s we operate the AOM near its bandwidth limit to achieve $\Delta \theta_i$ within $\sim$5~mrads range. We expect an enhanced range to $\sim$50~mrad using an AOD with smaller $v_{\rm s}$. 

For the small $\Delta \theta_i$ range in this work, we use a $D$=50~mm achromatic lens (Thorlabs AC508-750-B) with a focal length $F=750$~mm (II) to achieve a diffraction-limited focus with $w_{\rm G}=130~\mu$m at the grating interface. We use a 2400~line/mm holographic grating (III) with $d=0.42~\mu$m. Retro-diffraction efficiency close to $70\%$ is achieved by adjusting the incident polarization into the incident plane. Aided by a Faraday rotator, the AOM-double-passed beam is separated from the input with a polarization beamsplitter and is subsequently coupled into a single-mode polarization maintaining fiber as the output. The fiber coupling efficiency of $\sim 60\%$ is approximately $\omega_i-$independent within the limited $\Delta \theta_i$ range, and we achieve $(\tau_{\rm d})_{\rm max}\sim 100~$ps maximum delay (Fig.~\ref{fig:max(taud)}). The overall coupling efficiency $|\kappa|^2 \sim 0.05$ (Eq.~(\ref{equ:inOut})) is a result of non-ideal pulse picking, fiber coupling, and grating diffraction losses. We operate the shaper in the small signal regime, with single-pass AOM deflection efficiency of $|A_i|^2\sim 1-3\%$ to ensure the linear operation of the shaper up to $N=6$ pulses. This leads to an overall power efficiency of $\eta_{\rm P}\sim 10^{-5}$. Nevertheless, with ${\bf E}_{\rm in}$ at $P\approx 1~$kW peak power, ${\bf E}_{\rm out }$ at $\sim$10~mW peak power is still obtained at the output.

\subsection{Time-domain characterization}
\begin{figure}[htbp]
    \centering
    \includegraphics[width=0.45\textwidth]{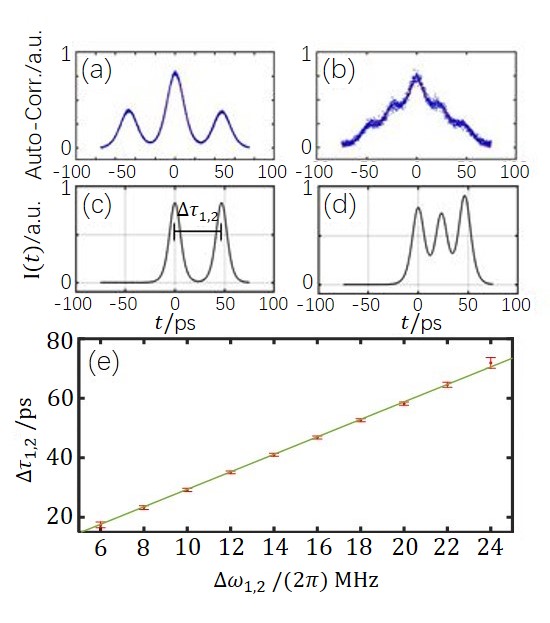}
    \caption{Auto-correlation measurements of the shaped pulses with $N=2$ sub-pulses associated with the two AOM frequencies $\omega_{1,2}=2\pi \times~76,~92$~MHz; and  $N=3$ sub-pulses, $\omega_{1,2,3}=2\pi\times~76,~84,~92$~MHz frequencies. Figs. (a)(b) are representative auto-correlation signals. Figs. (c)(d) give corresponding $I(t)$.  In Fig. (e), $\Delta \tau_{1,2}$ vs  $\Delta \omega_{1,2}$ extracted from ten measurements similar to those in Figs.~(a)(b) are plotted with red dots. The error bars represent the fit uncertainties. By fitting the data, the green line gives $\Delta \tau_{1,2}/{\rm ps}\approx 3.0 ~\Delta \omega_{1,2}/(2\pi\times {\rm MHz})$, in excellent agreement with Eq.~(\ref{equ:delay}). }
    \label{fig:2auto}
    \end{figure}

We verify the shaper performance by measuring both the time-domain $I(t)=|{\bf E}_{\rm out}(t)|^2$ and the frequency-domain $I(\omega)=|{\bf E}_{\rm out}(\omega)|^2$ of the shaped pulses under various $\{\omega_i, A_i^{\rm c},\varphi_i^{\rm c}\}$ controls. In the time domain, we use auto-correlation measurements based on phase-matched 2nd harmonic generation, with a APE PulseCheck 150 auto-correlator. Limited by the available auto-correlator range of delay, we only perform the measurement for $N=2$ and $N=3$ sub-pulse arrays. We keep the pulse picker fully open to enhance the signal levels. Typical results are given in Figs.~\ref{fig:2auto}(a)(b) for the cases of  $N=2,3$ sub-pulses respectively. With knowledge of sech$^2$-shaped single pulses, we de-convolve the auto-correlation curves to obtain the $\varphi_i$-averaged intensities $I(t)$ as in Figs.~\ref{fig:2auto}(c)(d).  

We repeat the $N=2$ sub-pulse measurements at different AOM deflection angles with $\Delta \theta_{1,2}$ controlled by the rf frequency difference $\Delta \omega_{1,2}=\omega_2-\omega_1$. In Fig.~\ref{fig:2auto}(e) the fitted relative delays $\Delta\tau_{1,2}=\tau_2-\tau_1$ are plotted vs $\Delta \omega_{1,2}$. The measured $\Delta\tau_{1,2}-\Delta\omega_{1,2}$ relation agrees excellently with Eq.~(\ref{equ:delay}) which predicts a slope within $0.5\%$ difference. The setup simplicity in Fig.~\ref{fig:setup}(a) allows us to extrapolate the high $\omega_i-\tau_i$ control quality from $N=2,3$ to shaped pulses with $N>3$ sub-pulses.

\subsection{Trap loss spectroscopy}
\begin{figure}[htbp]
\centering
\includegraphics[width=0.45\textwidth]{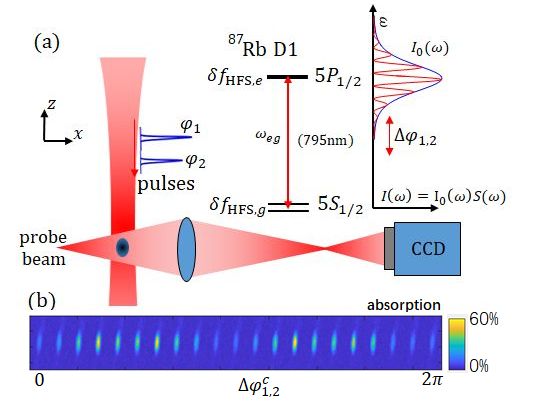}
\caption{Setup for ``super-resolved'' trap loss spectroscopy with shaped picosecond pulses. (a) Schematic diagram for shaped pulse excitation of $^{87}$Rb D1 line and the absorption imaging setup. Here $\delta f_{{\rm HFS},g}=6.8~$GHz and $\delta f_{{\rm HFS},e}=0.8~$GHz are hyperfine splittings. We take the shaped waveforms with $N=2$ sub-pulses and $\Delta \tau_{1,2}=96$~ps as an example. The spectrum given by $I(\omega)=I_0(\omega)S(\omega)$ displays $S(\omega)=\sin^2((\omega \Delta \tau_{1,2}+\Delta \varphi_{1,2})/2)$ interference. For the wide-band $I_0(\omega)$, the sub-pulses sequence is with a transform-limited frequency resolution $\delta \omega \approx  \pi/\Delta \tau_{1,2}$. For weak and repeated excitations, ``super-resolution'' features appear when $S(\omega)$ at $\omega=\omega_{eg}\pm \pi \delta f_{{\rm HFS},g}$ vanishes. Representative absorption images for the trap loss spectroscopy during a $\Delta\varphi_{1,2}=\varphi_2-\varphi_1$ scan over $4\pi$ (with $\Delta \varphi_{1,2}^{\rm c}$ scanning over 2$\pi$) are given in Fig. (b).
The data is from a denser set in Fig.~\ref{fig:TraplossFull}(b).}
\label{fig:atomsetup}
\end{figure}

We take the advantage of stable atomic frequency reference afforded by laser-cooled trapped alkali atoms, and use trap loss spectroscopy to characterise the waveforms of the shaped pulses in the frequency domain. The basic setup of the trap loss spectroscopy is sketched in Fig.~\ref{fig:atomsetup}. The central frequency $\omega_{\rm L}$ of the $\tau=11$~ps shaper input is adjusted to have (normalized) spectrum $I_0(\omega)$ with $\delta f_{\rm in}^{\rm L}\approx 30$~GHz width covering the $^{87}$Rb 5$S_{1/2}-5P_{1/2}$ D1 line (795~nm) centered at $\omega_{eg}=2\pi\times 377107~$GHz~\cite{SteckRb87}. 
The shaped ${\bf E}_{\rm out}$ pulses with spectrum $I(\omega)=S(\omega)I_0(\omega)$ are focused to overlap with an optically trapped $^{87}$Rb sample at $\sim 10~\mu$K temperature, prepared by laser cooling and moderate evaporation~\cite{YizunHe2019}. By subjecting the atoms to a sequence of shaped pulses with a repetition period $T_{\rm rep}\gg 1/\Gamma$, the D1 excitation followed by spontaneous emission during each repetition gradually heats the atoms up, leading to probabilistic escape of the hottest atoms whose kinetic energy is beyond the dipole trap depth $U$. Here $\Gamma=2\pi\times 5.7$~MHz is the D1 natural linewidth, and $S(\omega)=|s(\omega)|^2/(|s(\omega)|^2)_{\rm max}$ is a normalized pulse shaping spectrum factor, with $s(\omega)$ in Eq.~(\ref{equ:inOutf}) and $(|s(\omega)|^2)_{\rm max}=|\kappa \sum A_i^2|^2$.

We consider linearly polarized shaped pulses with duration $\tau_{\rm c}$ at typical $\sim 5$~mW peak power focused to a $\sim 50~\mu$m spot, with an $({\bf E}_{\rm out})_{\rm peak}\approx $40~kV/m peak field and an $\Omega_{\rm peak}= ({\bf E}_{\rm out})_{\rm peak} \cdot {\bf d}_{e g}/\hbar\approx 2\pi\times1$~GHz peak Rabi frequency. Here $|{\bf d}_{eg}|=1.46\times 10^{-29}$~C$\cdot$m is the D1 $\pi$-transition dipole matrix element~\cite{SteckRb87}, and 
we use $|g\rangle$, $|e\rangle$ to label the 5$S_{1/2}$ ground and 5$P_{1/2}$ excited states. The weak ${\bf E}_{\rm out}$ guarantees the pulse excitation probability $p_e=
|\int {\bf E}_{\rm out}(t)\cdot {\bf d}_{e g} e^{i \omega_{eg} t} d t/2\hbar|^2\ll 1$, which, as in Appendix \ref{AppendOptPump}, can be rewritten as $p_e=\Theta_0^2 S(\omega)/4$ with $\Theta_0\sim ({\bf E}_{\rm out})_{\rm peak}\cdot {\bf d}_{e g} \tau_{\rm c}/\hbar$ to be the maximum pulse area of the shaped waveform when all its sub-pulses constructively contribute to the excitation in this 2-level model.

Experimentally we adjust the trap depth $U\approx k_B\times 60~\mu$K, the ``heating'' time $T_{\rm total}$ and $N_{\rm total}=T_{\rm total}/T_{\rm rep}$, so as to achieve a high sensitivity for the fractional trap loss $l=\Delta N_{\rm a}/N_{\rm a}$ as a function of $p_e$ during sub-pulse phase scans. Here $N_{\rm total}$ represents the number of the shaped pulses the atoms are exposed to, $N_{\rm a}$ represents the estimated atoms number in the trap and $\Delta N_{\rm a}$ represents the reduced atoms number. Assuming kinetic energy gain per excitation-emission cycle to be twice the recoil energy~\cite{SteckRb87} ($\Delta E\approx k_B \times 700~$nK), for $N_{\rm total}p_e \sim 100$ the cumulative kinetic energy gain would be comparable to trap depth $U$, leading to substantial trap loss. The functional form of $l(p_e)$ depends on the trapping potential details and is therefore not precisely known. However, simple analysis suggests $l$ increases monotonically with $p_e$ and is more $p_e-$sensitive for smaller $l$. With a detailed study of $l(p_e)$ left for a future publication, here we use the trap loss $l$ to monitor $p_e$ subjected to shaped pulses in repeated experiments.

Beyond the 2-level model, the experimentally estimated $p_e$ is in addition compared with a 3-level model detailed in Appendix \ref{AppendOptPump}, which by accounting for hyperfine optical pumping~\cite{OpticalPumpingBook} leads to $p_e=\Theta_0^2 \tilde S(\omega_{eg},\delta f_{{\rm HFS},g})/4$. Here the spectral response function $\tilde S(\omega,\delta f)=\frac{2 S(\omega+\pi\delta f)S(\omega-\pi\delta f)}{S(\omega+\pi\delta f)+S(\omega-\pi\delta f)}$ may vary sharply over $\delta f=\delta f_{{\rm HFS},g}$, even for shaped waveforms with an overall duration $\tau_{\rm c}\ll 1/\delta f_{{\rm HFS},g}$ such that the associated transform-limited frequency resolution $1/2\tau_c$ is much larger than $\delta f_{{\rm HFS},g}$. This frequency ``super-resolution'' effect is related to population redistribution when the spectrum factor $S(\omega)$ vanishes near a hyperfine resonance. 
Observation of such ``super-resolved'' features requires a shaper with substantially higher precision and stability than those required for observing simple 2-level dynamics.

\begin{figure}[htbp]
\centering
\includegraphics[width=0.45\textwidth]{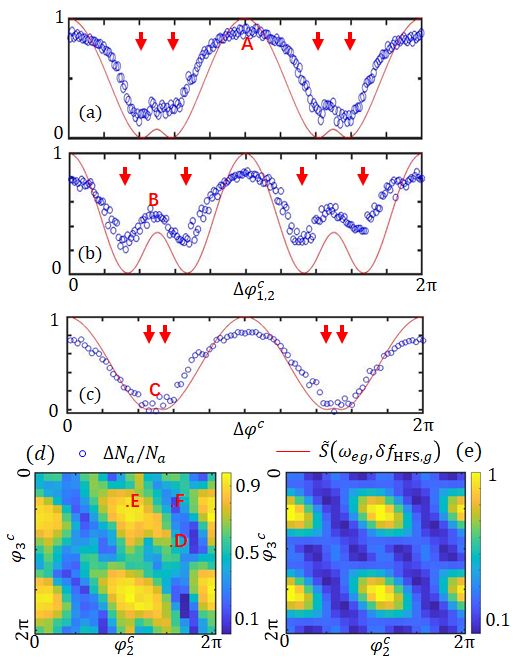}
\caption{Phase-scanning trap loss spectroscopy for shaped waveforms with $N=2$ (a,b), $N=6$ (c), and $N=3$ (d) sub-pulses. The inter sub-pulse delays are $\Delta\tau_{1,2}=24$~ps, 96~ps, 12~ps and 24~ps from (a) to (d). See main text for detailed descriptions.}
\label{fig:TraplossFull}
\end{figure}

We start with the simplest pulse shaping with $N=2$ sub-pulses at various delays $\Delta \tau_{1,2}$.  For the 2-level model, the double resonant pulses should lead to $p_e\propto S(\omega_{eg})= \cos^2((\omega_{eg} \Delta \tau_{1,2} +\Delta \varphi_{1,2})/2)$ to dictate the trap losses (Fig.~\ref{fig:atomsetup}). Experimentally, we set $A_1\approx A_2$ by adjusting the control rf amplitudes, and then scan $\Delta \varphi_{1,2}^{\rm c}=\varphi_1^{\rm c}-\varphi_2^{\rm c}$ in repeated experiments to record the corresponding trap losses. According to Eq.~(\ref{equ:inOut}) we have $\Delta \varphi_{1,2}= 2 \Delta \varphi_{1,2}^{\rm c}$ up to a constant offset. By adjusting the offset, the trap losses $l=\Delta N_{\rm a}/N_{\rm a}$ vs $\Delta \varphi_{1,2}^{\rm c}$ are plotted in-phase with $S(\omega_{eg})\sim\cos^2(\Delta \varphi^{\rm c}_{1,2})$ in Fig.~\ref{fig:atomsetup}.  ``Sub-$\pi$'' deviation of data from the sinusoidal form are found near the trap loss minimum, which is more pronounced in Fig.~\ref{fig:TraplossFull}(b) with the longer $\Delta \tau_{1,2}=96$~ps, but is also seen in Fig.~\ref{fig:TraplossFull}(a) with $\Delta \tau_{1,2}=24$~ps. Under both delays, locations for the curved deviations are captured by the spectral response function $\tilde S(\omega_{eg},\delta f_{{\rm HFS},g})$. In particular, the trap loss minimum (marked with red arrows) are reached near $(\Delta \varphi_{1,2})_m=2\pi(1\pm \delta f_{{\rm HFS},g}\Delta \tau_{1,2}/4)$ such that the shaped pulse spectrum factor $S(\omega)$ vanishes at the $\omega=\omega_{eg} \pm \pi\delta f_{{\rm HFS},g}$ hyperfine resonances.
The ``sub-$\pi$'' features in Figs.~\ref{fig:TraplossFull}(a)(b) thus correspond to resolution beyond the $1/2\tau_c$ transform limit in the frequency domain.

Similar ``sub-$\pi$'' features are also found in $N=3$ phase-scanning spectroscopy in Fig.~\ref{fig:TraplossFull}(d). Here we again set approximately equal $A_{1,2,3}$. With $\Delta\tau_{1,2}=\Delta\tau_{2,3}=24$~ps inter-pulse delays and the $\tau_{\rm c}\approx \Delta \tau_{1,3}+\tau=59$~ps duration, we scan $\varphi_{2,3}^{\rm c}$ relative to $\varphi_1^{\rm c}$ in 2D, with $\varphi_1^{\rm c}$ fixed in repeated experiments. The 2D trap losses again demonstrates features beyond the expected excitation probability by the simple 2-level model which suggests $p_e\propto S(\omega_{eg})= \big(\sin^2(\Delta\varphi^{\rm c}_{1,2})+\sin^2(\Delta\varphi^{\rm c}_{2,3})+\sin^2(\Delta\varphi^{\rm c}_{3,1})\big)/3$. The ``sub-$\pi$'' features  again appear in the small $l$ regime and are well-captured by the corresponding $\tilde S(\omega_{eg},\delta f_{{\rm HFS},g})$ (Appendix \ref{AppendOptPump}) in Fig.~\ref{fig:TraplossFull}(e).

We finally demonstrate pulse shaping with up to $N=6$ sub-pulses (Fig.~\ref{fig:TraplossFull}(c)). Here, with $\Delta\tau_{i,i+1}=12$~ps equal inter-pulse delay and approximately equal $A_i^2$, we uniformly vary $\varphi^{\rm c}_{1,3,5}=\varphi^{\rm c}_{2,4,6}+\Delta \varphi^{\rm c}$ of the $\tau_{\rm c}\approx 83$~ps waveform and to record trap loss $l$ in repeated experiments. Here $\varphi^{\rm c}_{1,3,5}$ is set according to the marker ``${\bf E}$'' in Fig.~\ref{fig:TraplossFull}(d) for 3-pulse full constructive interference to the atomic excitation. By adjusting $\varphi^{\rm c}_{1,3,5}$ and $\varphi^{\rm c}_{2,4,6}$ out of phase, however, we see substantially reduced trap loss $l$ near $\Delta\varphi^{\rm c}=\pi/2,3\pi/2$. Comparing with Fig.~\ref{fig:TraplossFull}(a)(b), the ``super-resolution features'' at small $l$ nearly disappear, which is quite expected since with $\Delta \tau_{i+1,i}=$12~ps the spectrum factor $S(\omega)$ varies so slowly that little difference is found across the $\delta f_{{\rm HFS},g}$ interval. Correspondingly, the spectral response function $\tilde S(\omega_{eg},\delta f_{{\rm HFS},g})$ is largely sinusoidal (though near $\Delta \varphi^{\rm c}=\pi/2,3\pi/2$ the sub-$\pi$ features do still exist, as marked. The expected feature is below our experimentally achieved signal-to-noise).

The contrast of the experimentally measured $l=\Delta N_{\rm a}/N_{\rm a}$ as in Figs.~\ref{fig:TraplossFull}(a-d) are generally less than $100\%$ during the phase scans, which is partly explained by the amplitude-imbalanced sub-pulses with reduced $S(\omega)$ contrast in the first place. We emphasize that the phase-scanning data are from $\sim6$ hours of measurements where the cold atom sequence cycles every 1.5 seconds. We do not observe phase-sensitive fluctuation of trap loss data, suggesting the phase stability of the shaper during the measurement repetitions. The phase stability is passively maintained over several days in our temperature-stabilized lab. 

\begin{figure}[t]
\centering
\includegraphics[width=0.5\textwidth]{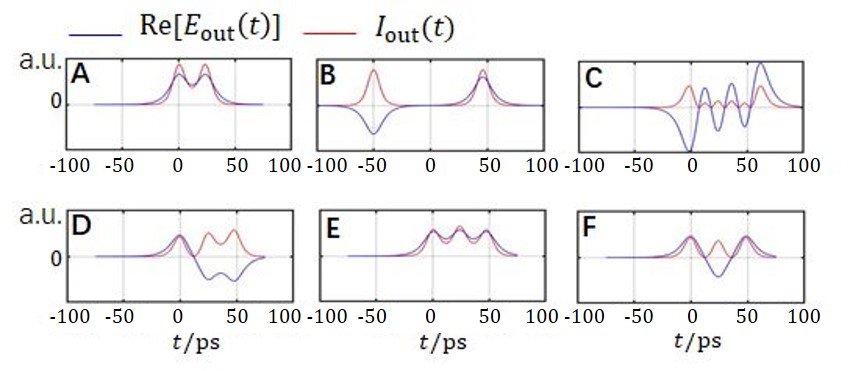}
\caption{Reconstruction of shaped waveform intensity $I_{\rm out}(t)$ and ${\rm Re}[ E_{\rm out}(t)]$ quadrature (in the $\omega_{eg}$ rotating frame) according to Eq.~(\ref{equ:inOut}), for $N-$ sub-pulse array with uniform $\{A_i\}$ and specific phase combination $\{\varphi_i\}$ marked in Figs.~\ref{fig:TraplossFull}. 
} 
\label{fig:simulation}
\end{figure}

To illustrate the actual optical waveform obtained from the shaper for the trap loss spectroscopy,  we calculate the complex ${\bf E}_{\rm out}(t)$ using Eq.~(\ref{equ:inOut}) with the transform-limited ${\bf E}_{\rm in}(t)$. Typical reconstructed waveforms are presented in Fig~\ref{fig:simulation}.

\section{Discussions}\label{Sec:Discucssion}

\begin{figure}[htbp]
\centering
\includegraphics[width=0.45\textwidth]{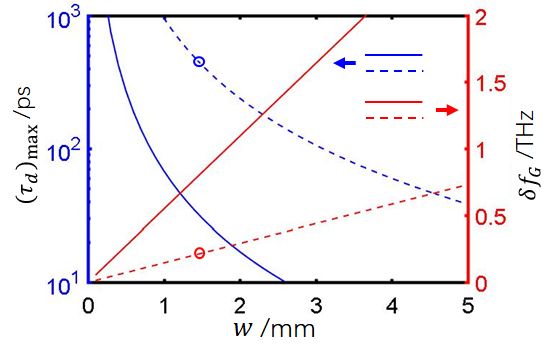}
\caption{Typical $(\tau_{\rm d})_{\rm max}$ (associated with $z_{\rm R,G}=\lambda F^2/\pi w^2$) and $\delta f_{\rm G}$ (associated with $w_{\rm G}= \lambda F/\pi w$) in the shaper scheme vs the Gaussian waist $w$ of the input laser near the AOD, for pulse shaper with various lens focal length $F$ ($F=200$~mm, solid line; $F=750$~mm, dash line). We consider $\lambda=0.8$~$\mu$m and a 2400~line/mm grating with grating constant $d=0.42$~$\mu$m. The colored circles mark the limits in this work. With $\delta f_{\rm G}$ we can estimate the delay resolution $(\tau_{\rm d})_{\rm min}=\pi/(2\delta f_{\rm G})$. For operation of the shaper in the ``broadband regime'' of grating diffraction, $\delta f_{\rm G}$ sets the upper bound for the input pulse bandwidth $\delta f_{\rm in}^{\rm L}$ which is also equal to the modulation bandwidth $\delta f_{\rm M}$ of the output waveforms with duration $\tau_{\rm c}\leq (\tau_{\rm d})_{\rm max}+\tau$.}
\label{fig:max(taud)}
\end{figure}

\subsection{Maximum/minimum delay, and waveform modulation bandwidth}\label{SecMDMLimit}

To understand the delay generation based on retro-diffraction (Fig.~\ref{fig:setup}(a)), we consider the grating diffraction in a ``broadband regime'' with single mode diffraction bandwidth $\delta f_{\rm G}= \sqrt{4d^2/\lambda^2-1}c /(\pi w_{\rm G} )$ much larger than the spectrum width of the incident pulse $\delta f_{\rm in}^{\rm L}$. Here $\delta f_{\rm G}$ is set by the duration of the pulse-grating interaction for each incoming beam with diffraction limited waist $w_{\rm G}= \lambda F/\pi w$, and we have focal length $F$, incoming beam waist $w$, grating constant $d$ and wavelength $\lambda$ as in Fig.~\ref{fig:setup}(a). In this $\delta f_{\rm G}\gg \delta f_{\rm in}^{\rm L}$ regime, the retro-diffraction only contributes to the loss factor $\kappa$ as in Eq.~(\ref{equ:inOut}) without affecting spectral phase of each $E_{i,{\rm out}}$. The maximum delay $(\tau_{\rm d})_{\rm max}$ is limited by the retro-diffraction efficiency within the Rayleigh range $z_{\rm R,G}=\pi w_{\rm G}^2/\lambda$ of the focused Gaussian beam as $(\tau_{\rm d})_{\rm max}=2 z_{\rm R,G}/c=2 F^2\lambda/(\pi w^2 c)$. 

The minimal delay $(\tau_{\rm d})_{\rm min}$ between sub-pulses is instead set by the AOD frequency resolution $\Delta \omega=2 v_{\rm s}/w$  through Eq.~(\ref{equ:delay}) as $(\tau_{\rm d})_{\rm min}=2 \lambda F/(\pi w c\sqrt{4d^2/\lambda^2-1}$). We thus find $(\tau_{\rm d})_{\rm min}=\pi/(2\delta f_{\rm G})$. For transform-limited sech$^2$ input with duration $\tau$ and $\delta f_{\rm in}^{\rm L}\approx 0.315/\tau$, $\delta f_{\rm in}^{\rm L} \ll \delta f_{\rm G}$ as in this work thus suggests $(\tau_{\rm d})_{\rm min}\ll \tau$ for arbitrarily shaping of the output. Therefore, the shaper can produce arbitrary waveforms with duration $\tau_{\rm c}\leq (\tau_{\rm d})_{\rm max}+\tau$ within the modulation bandwidth $\delta f_{\rm M}=\delta f_{\rm in}^{\rm L}$ set by the input pulse bandwidth. 

To give concrete examples, in Fig.~\ref{fig:max(taud)} we plot the $w_{\rm G}-$limited $\delta f_{\rm G}$ and $z_{\rm R,G}$-limited $(\tau_{\rm d})_{\rm max}$ as a function of input beam size $w$ for typical experimental settings. Here the $\delta f_{\rm G}$ lines set the $\delta f_{\rm M}=\delta f_{\rm in}^{\rm L}$ limit for operation of the shaper in the ``broadband regime''. The limits for $F=750$~mm, $w=1.5$~mm as in this work is also marked. Operation of the shaper with $(\tau_{\rm d})_{\rm max}$ in {\it nanosecond regime} is possible by increasing the ratio $F/w$ at moderate reduction of $\delta f_{\rm G}$.

The shaper properties are more complicated in the ``narrow-band regime'' of diffraction with $\delta f_{\rm G} \ll \delta f_{\rm in}^{\rm L}$. In this case the diffraction selects $E_{i,{\rm out}}$ within a $\delta f_{\rm G}$ bandwidth, which itself depends on the incident beam size that varies across the Rayleigh range $z_{\rm R,G}$. The delay-dependent filtering leads to output waveforms beyond Eq.~(\ref{equ:inOut}). Also, the output cannot be ``arbitrarily'' programmed particularly since $(\tau_{\rm d})_{\rm min}$ is close to the pulse duration at the output. Understanding and optimization of the shaper in this ``narrow-band regime'', particularly for ultrafast pulse shaping, is an interesting topic for future study. 

\subsection{Extended delay and bandwidth}\label{SecExtension}
Beyond the basic Fig.~\ref{fig:setup}(a) setup, enhancement of $(\tau_{\rm d})_{\rm max}$ beyond nanoseconds is possible by individually managing groups of sub-pulses with 4-F imaging systems. Each sub-pulse group is then optimally retro-diffracted at different focal $w_{\rm G}$ location. Alternatively, the large area gratings can be replaced with an integrated array of curved micro-mirrors for wavefront-matched retro-reflections over extended delaying distances. The replacement of gratings with mirrors would compromise the delay tunability. However, continuous tunability of inter-pulse delays is often not crucial, as long as the pulse spacing is short comparing with the timescales for the dynamics of interest, as in nonlinear spectroscopy~\cite{MDCSAnnualReview2015} and quantum controls~\cite{GenovFieldError2014,Chuang2016}. With the extension and by keeping $\delta f_{\rm M}=\delta f_{\rm in}^{\rm L}$, our shaper scheme would support highly complex waveforms with long duration $\tau_{\rm c}\leq (\tau_{\rm d})_{\rm max}+\tau$ and extremely large $\delta f_{\rm M}\times \tau_{\rm c}$ product.

\subsection{Phase and amplitude stabilization}\label{SecWFStability}

The passive waveform stability demonstrated in this work is nevertheless subjected to long-term drifts of optical alignments, as suggested by the Fig.~\ref{fig:TraplossFull} data. In addition, the phase stability is expected to degrade when sophisticated $(\tau_{\rm d})_{\rm max}$-extending beam-steering optics are introduced. Fortunately, it should be quite straightforward to actively stabilize the multiple delay lines. For example, a frequency-stabilized CW laser at a wavelength close to that for the pulse laser can be injected to co-propagate with the sub-pulses. We can then monitor the amplitudes and phases of the CW beams double-deflected by the $\omega_i$-sound waves, with heterodying measurements at the output to resolve the $2\omega_i$ beat signals. As such, the same $\{A_i,\varphi_i\}$ pulse control parameters are retrieved and stabilized by adjusting $\{A_i^{\rm c},\varphi_i^{\rm c}\}$ for the rapid AOD sound-wave updates. To avoid cross-talks between the pulsed shaper outputs and the CW monitoring laser, the CW measurements can be operated stroboscopically in between the shaped pulses.

\subsection{Reaching the efficiency limit}\label{SecEffLimit}
The programmable pulse shaper relies on rf-control of the $\{A_i,\varphi_i,\tau_i\}$ sub-pulses as prescribed by Eq.~(\ref{equ:inOut}). The programming is straightforward in the small signal regime, with $A_i\propto A_i^{\rm c}$ and $\varphi_i=2\varphi_i^{\rm c}$ up to a phase offset. However, operation of the shaper in this regime quite unavoidably leads to very low efficiency (Sec.~\ref{Sec:Exp}), with an average sub-pulse AOD deflection efficiency $\langle |A_i|^4 \rangle \sim 10^{-4}$ and overall power efficiency $\eta_{\rm P}\sim 10^{-5}$ in this work. To apply the shaper to more power-demanding experiments such as for error-resilient quantum controls of large samples, the shaper efficiency needs to be substantially improved, ideally to the fundamental $\eta_{\rm P}\sim 1/N^2$ limit for the filtering-based arbitrary shaper. The AOD nonlinearity can be partly suppressed with better amplification linearity at high output rf power, and more efficient acoustic-optical transduction. However, ultimately the nonlinearity in the  $A_i=A_i(\{A_j^{\rm c},\varphi_j^{\rm c}\})$, $\varphi_i=\varphi_i(\{A_j^{\rm c},\varphi_j^{\rm c}\})$ controls would emerge due to depletion of the 0$^{\rm th}$-order beams during Bragg-diffractions. Therefore, to reliably operate the pulse shaper at the $\eta_{\rm P}\sim 1/N^2$ efficiency limit, we expect global $\{A_j^{\rm c},\varphi_j^{\rm c},\omega_j\}$-optimization to compensate for the nonlinear responses including cross-talks, so as to achieve the target $\{A_i,\varphi_i,\tau_i\}$ control. Given the stability of the nonlinear effects to be compensated for, we expect efficient optimization based on in situ CW laser measurements (Sec.~\ref{SecWFStability}) for reliable pulse shaper operation near the efficiency limit.

\subsection{Comparison to other shapers}
As mentioned, our method is a form of acoustic-optical programmable filter that supports delays much longer than those limited by the crystal size in traditional AOPDF~\cite{AOTFWeinerOL1993, AODispersiveFilter1997}. The AOD acts as a multi-port beamsplitter that interferometrically couples programmable delay lines into a single mode output. Similar multi-port interferometry using multiple optical beamsplitters can also generate the delayed pulse sequences using individually stabilized delay lines~\cite{InterferometerSiders1998,Dromey2007,2DSpecbyAO2007,stableInstruments2009} with partial or full phase/amplitude programmability. Comparing with the mechanical delay-line based pulse sequence generators, here the number of delay lines $N$ and the amount of delays $\{\tau_i\}$ are rapidly re-configurable, an important advantage for optimal quantum control. In addition, the parallel propagation of multiple delay lines through a same set of optics (Fig.~\ref{fig:setup}a) ensures vibration-insensitive passive phase stability. The relative phases can in addition be conveniently stabilized by a single CW laser measurement (Sec.~\ref{SecWFStability}). 

In the following we discuss the connection between our method with the Fourier transform pulse shaping (FTPS~\cite{FTPS1983,Weiner2000SLMreview,fspsStobrawa2001,CDMAWeiner1998}), and the direct space-to-time pulse shaping (DSTPS~\cite{OriginOfDSTPS1992,WeinerDST1999, diffractiveAWGOptLett2010, CompactDST2011}). 

In typical FTPS setups (Fig.~\ref{fig:setup}(b)), the input pulses ${\bf E}(t)$ are dispersed to establish a $f-y$ frequency-position mapping. Programmable spectral phase and transmission are applied as $A(f(y))e^{i\varphi(f(y))} {\bf E}(f(y))$, before another diffraction recombining the frequency components into the output. FTPS has been very useful in ultrafast science, particularly for optimal control~\cite{controlbyPS1998,nonlinearRamanSpec2002} and multi-dimensional spectroscopy~\cite{MDCSRCLScienceWarren2003,2DftSpec2009}. However, FTPS appears not particularly suitable for shaping long pulses for narrow-band applications. In particular, as discussed in the Motivation section, precise control of optical transition prefers long control time $\tau_{\rm c}$ with nearly resonant pulses. To stretch narrow-band pulses with long optical delays, highly dispersive diffractions needs to be stably maintained in FTPS. Furthermore, the transmission function $A(f(y))e^{i\varphi(f(y))}$ needs to be rapidly modulated in space, usually leading to unwanted diffraction losses in the single mode outputs~\cite{STCoupling2008,STcoupling2009}. Such spatial-temporal coupling effects can be mitigated if the spectrum of the shaped pulse forms stable and sparse combs. Indeed, more recent developments allow FTPS to engineer high repetition frequency combs for completely controlled optical arbitrary waveform generation (OAWG)~\cite{Weinerlinebyline2005,Weinerawg2008,WeinerLinebyLine2011,890MHzWeiner2012}. However, it still appears difficult to isolate individual shaped pulses from existing OAWGs for e.g., combined maneuvers in quantum controls such as state preparation and detection.

In typical DSTPS setups (Fig.~\ref{fig:setup}(c)), ${\bf E}(x,y,z,t)$ propagating along ${\bf e}_x$ is spatially modulated across its wavefront along ${\bf e}_y$. With the retro-diffractive optics similar to this work, the spatial location $y$ determines optical delay $t$ through a $t-y$ mapping. The delay-dependent transmission is directly programmed into the output as  $\sum_y A(t(y))e^{i\varphi(t(y))} {\bf E}(t(y))$. By operating the shaper in the time domain, the transmission $A(t(y))e^{i\varphi(t(y))}$ can be a smooth function even for long optical delays. Therefore, comparing with FTPS, DSTPS is more suitable to stretch a single narrow band pulse into a longer and phase programmable pulse without being severely affected by the spatial-temporal coupling losses. However, the spatial wavefront division generally affects the single-mode quality. The mapping also limits the shaping efficiency, particularly for generating sparse and isolated pulses since a majority of the pre-aligned wavefronts needs to be blocked.

The basic idea of the multi-delay based shaping scheme (Fig.~\ref{fig:setup}(a)) follows DSTPS as to directly program the waveforms in the time domain. However, here the wavefront is divided in ${\bf k}-$space, and the shaping effect as in Eq.~(\ref{equ:inOut}) is obtained by instead a $t-{\bf k}$ mapping.  The ${\bf k}$-vectors are conserved quantities before and after the division. Therefore, single-mode wavefront division in ${\bf k}-$space is generally more precise and stable than cutting wavefronts in real space. Furthermore, the AOD double-pass deflection combined with beam waist managements ensures uniform output coupling efficiency insensitive to optical delays, effectively nullifying the cross-talk between the delay and amplitude/phase of isolated sub-pulses. Cross-talks among sub-pulses with short relative delay close to $(\tau_{\rm d})_{\rm min}$ (Sec.~\ref{SecMDMLimit}) can also be corrected in ways similar to overcoming the nonlinear sub-pulse cross-talks (Sec.~\ref{SecEffLimit}). Finally, the application of AOD facilitates fast updates of the shaped optical waveforms within microseconds, an important advantage for various feedback controls.

\subsection{Summary and outlook}
In this work we have proposed a simple method that precisely shapes transform-limited picosecond pulses into sub-THz optical waveforms. We have provided proof-of-principle demonstration of the scheme, by shaping $\sim$11~ps pulses into arrays of sub-pulses, resulting in single-mode outputs with up to 100~ps duration and $\sim 30$~GHz bandwidth. The precision, stability, and nearly arbitrary programability of the waveforms are corroborated with precise measurements. In particular, by operating the shaper in a small signal regime, we observe GHz-level ``super-resolved'' features of atomic spectroscopy with laser-cooled $^{87}$Rb atoms. We have outlined methods for extension beyond this demonstration, for generation of sub-THz waveforms with duration beyond nanoseconds, with precise waveforms stabilized by active feedbacks.

Toward error-resilient quantum control of optical electric dipoles, the shaper needs to be operated beyond the small signal regime so as to achieve the $\eta_{\rm P}\sim 1/N^2$ power efficiency limit. For the purpose, during the waveform stabilization (Sec.~\ref{SecWFStability}) one may optimize $A_i(\{A_j^{\rm c},\varphi_j^{\rm c}\})$ and $\varphi_i(\{A_j^{\rm c},\varphi_j^{\rm c}\})$ toward target values during efficiency-limited AOD deflections (Sec.~\ref{SecEffLimit}). With  $\eta_{\rm P}\sim 1/N^2$ and by using a same picosecond laser as in this work, we expect sub-pulses with $\sim10$~W peak power for an $N=5$ array, which, if being focused to a $\sim 20~\mu$m laser spot, would lead to $^{87}$Rb D1-coupling Rabi-frequency of $\Omega_{\rm peak}\sim 2\pi\times100$ GHz and $\Theta_i\sim \pi$ pulse area.  Together with a moderate enhancement of the control time $\tau_{\rm c}=(\tau_{\rm d})_{\rm max}$ (such as by using AOD with a lower speed of sound $v_{\rm s}$), the $N=5$ array of isolated sub-pulses would already enable composite pulse control in ways similar to those in NMR research~\cite{Odedra2012,GenovFieldError2014,Chuang2016}, for  emergent applications in quantum optics~\cite{scully2015, YizunHe2019}. Even more complex waveforms~\cite{Navin2005grape,NatCommDu2015} for high-fidelity single qubit gates on a strong optical transition may be achieved with a better focused laser beam together with longer $\tau_{\rm c}$ and larger $N$. This or similar techniques would also strengthen the prospects of atomic state control and measurements with a mode-locked laser~\cite{Freegarde1995,Immanuel1997,Jayich2014,campbell2019}, including ultra-fast control of atomic motion for precision measurements ~\cite{PritchardAIreview2009,Interferometry2014} and quantum information processing~\cite{Monroe2014,Monroe2017}. 
Taking one more step further,  by combining multi-colored waveforms from multiple shapers, we anticipate multi-level control of atoms for accurate preparation of specific electronic quantum states in large samples with high optical depths, such as those with maximum multi-photon coherence, for emergent nonlinear and quantum optical applications.

\section*{Funding Information}
National Key Research Program of China (2017YFA0304204, 2016YFA0302000); 
National Science Foundation of China (11574053). Shanghai Scientific Research Program (15ZR1403200).

\section*{Disclosures}
Techniques assoicated with the pulse shaping scheme is under Chinese Patent Application No. 201911401004.X, Unpublished (filing date Dec. 31, 2019).

\section*{Acknowledgments}
We are grateful to R.~Brown and G.~Bruce for valuable suggestions to the manuscript, to Z.-S.~Tao and C.-S.~Tian for helpful discussions. We thank CIQTEK for support by providing a high-resolution arbitrary sequence generator. 

\appendix

\section{Trap loss spectroscopy}\label{AppendTrapLoss}
This appendix provides additional details on trap loss measurements and the associated theoretical models. 

\subsection{Setup and measurement procedures}\label{AppendSetup}

The laser-cooling part of the experimental setup is similar to that in ref.~\cite{YizunHe2019}: Up to 10$^7$ $^{87}$Rb atoms are loaded into a magneto-optical trap in less than 1 sec.  Assisted by polarization gradient cooling, up to 10$^5$ atoms are then loaded into a 1064~nm crossed optical  dipole  trap  at $\sim$0.5~mK initial depth,  which  are subsequently transferred to a hybrid trap composed of both the 1064~nm trap and a 840~nm dimple trap. This system is designed for evaporation of the sample to quantum degeneracy~\cite{DimpleDalibard2011}. In this work, slight evaporation by reducing the hybrid trap depth produces the $\sim10~\mu$K atomic sample with up to $N_{\rm a}=10^4$ atoms. We adjust the power ratio of the dipole and dimple traps to achieve nearly spherical atomic sample with characteristic Gaussian radius of $\sigma \approx 10~\mu$m. 

To estimate the dipole trap depth, we perform a series of absorption imaging measurements on the D2 line  at various probe detuning. Comparing with free atoms, the $5S_{1/2}, F=2- 5P_{3/2}, F'=3$ resonance for the dipole-trapped atoms is shifted in the absorption spectrum by a MHz-level light-shift $\Delta$. We adjust the dipole trap laser intensities to have $\Delta\approx 2\pi\times 2$~MHz. Aided by knowledge of $5S_{1/2}$ and $5P_{3/2}$ polarizabilities, we estimate $U\approx h\times 1.3$~MHz and thus $U=k_B\times 60~\mu$K trap depth with $k_B$ the Boltzmann constant. 

We expose the trapped atoms to a shaped pulse train for a duration $T_{\rm total}$ and $N_{\rm total}=T_{\rm total}/T_{\rm rep}$ pulses. For weak excitation with atoms being excited probabilistically between long intervals, each excitation leads to twice the photon recoil energy transfer with $\Delta T =\frac{\hbar k^2}{m k_B}=700~$nK~\cite{SteckRb87} ($m$ is the mass of $^{87}$Rb atom). After the heating pulses, the atoms remaining in the trap are hold for an additional time of 30~ms, before being released for absorption measurements. We allow a 100~$\mu$s free-flight time for the sample to expand from $\sigma\sim 10~\mu$m to about $20~\mu$m in size. Absorption of a resonant D2 probe beam by the atoms is then imaged on a CCD camera as $I_1$. Absolute absorption $A=\Delta I/I$ is estimated by taking an reference image $I_2$ in absence of the sample (An array of 21 processed absorption images are given in Fig.~\ref{fig:atomsetup}b.). With knowledge of atomic absorption cross-section~\cite{SteckRb87}, we then estimate the remaining atom number $N_{\rm r}$ and thus $\Delta N_{\rm a}=N_{\rm a}-N_{\rm r}$ and $l=\Delta N_{\rm a}/N_{\rm a}$. Trap loss measurements at certain $\{A_i^{\rm c},\varphi_i^{\rm c},\omega_i\}$-control combinations are repeated for 4-6 times. The absorption images are individually processed to estimate both the average fractional loss $\Delta N_{\rm a}/N_{\rm a}$ and its standard deviation.  We empirically set $T_{\rm total}=20\sim50$~ms ($T_{\rm rep}=250$~ns, $N_{\rm total}= 8\sim 20 \times 10^4$) for sufficient heating and for high contrast $l=\Delta N_{\rm a}/N_{\rm a}$  during the $\{\varphi_i^{\rm c}\}$-scans. 

\subsection{``Super-resolved'' atomic frequency response}\label{AppendOptPump}

The D1 line of $^{87}$Rb as in Fig.~\ref{fig:atomsetup} is composed of 4 transitions separated by $\delta f_{{\rm HFS},g}=6.8$~GHz and $\delta f_{{\rm HFS}, e}=0.8$~GHz hyperfine splittings. Although features of trap loss spectroscopy is largely captured by the 2-level picture discussed in the main text, the simple model is incapable of revealing ``super-resolved'' features as in Figs.~\ref{fig:TraplossFull}, which, as will be clarified in this sub-section, are associated with hyperfine optical pumping effects~\cite{OpticalPumpingBook}. Here we illustrate the essential physics with a simple 3-level model. The validity of the conclusions is confirmed by numerical simulation of the full D1 line excitation dynamics. 

We consider linearly polarized shaped pulses with electric field  ${\bf E}_{\rm out}(t) + c.c.$ composed of an array of $\tau=11~$ps sub-pulses and up to $\tau_{\rm c}=100$~ps duration, with central frequency $\omega_{\rm L}$ to drive the $\pi$-transition of the $^{87}$Rb D1 line. The 5P$_{1/2}$ hyperfine splitting $f_{{\rm HFS},e}\ll 1/\tau_{\rm c}$ and is ignored. The light-atom interaction is effectively modeled by a simple 3-level Hamiltonian under rotating wave approximation as
\begin{equation}
\begin{array}{l}
    H_{\rm eff}=\hbar \pi\delta f_{{\rm HFS},g}(|b\rangle\langle b|-|a\rangle\langle a|)+\hbar (\Delta_e-i\Gamma/2) |e\rangle\langle e|+\\
    ~~~~~~~~~~~~\frac{\hbar \Omega_a(t)}{2}|e\rangle\langle a|+\frac{\hbar \Omega_b(t)}{2}|e\rangle\langle b|+h.c.
\end{array}
\end{equation}
Here $|a\rangle$, $|b\rangle$, $|e\rangle$ represent a particular set of Zeeman sublevels with $m_F$ magnetic quantum number in the $5S_{1/2} F=1$, $5S_{1/2} F=2$ and $5P_{1/2} F'=1,2$ levels respectively, and $\Delta_e=\omega_{eg}-\omega_{\rm L}$ and $\omega_{eg}$ is the mean transition frequency of the four hyperfine lines.  With the mode-locked laser central frequency tuned to $\omega_{\rm L}=\omega_{eg}$ (Fig.~\ref{fig:atomsetup}), we effectively set the central detuning $\Delta_e=0$. The laser coupling Rabi frequencies are given by $\Omega_a(t)={\bf E}_{\rm out}(t)\cdot {\bf d}_{a e}e^{i\omega_{eg} t}/\hbar$ and similarly $\Omega_b(t)={\bf E}_{\rm out}(t)\cdot {\bf d}_{b e}e^{i\omega_{eg} t}/\hbar$. The pulsed excitation and decay dynamics can be evaluated with single-atom density matrix $\rho(t)$, which is governed by the master equation,
\begin{equation}
    \dot{\rho}=\frac{1}{i \hbar}(H_{\rm eff}\rho-\rho H_{\rm eff}^{\dagger})+C_a \rho C_a^{\dagger}+C_b \rho C_b^{\dagger}.\label{equMaster}
\end{equation}
Here $C_a=\sqrt{\Gamma_a}|a\rangle\langle e|$ and $C_b=\sqrt{\Gamma_b}|b\rangle\langle e|$ are quantum jump operators for effective population recycling in the 3-level model. We thus set total spontaneous emission rate $\Gamma=\Gamma_a+\Gamma_b$ for the self-consistent modeling. With $\Gamma\tau_{\rm c}\ll1$, the spontaneous decay is negligible during each pulsed excitation. On the other hand, with $T_{\rm rep}=250$~ns~$\gg 1/\Gamma$, spontaneous emission resets population back to the ground states before the next excitation.  

It is important to notice that with the negligible $\delta f_{{\rm HFS},e}$, hyperfine Raman-coupling associated with $\Omega_a \Omega_b^*$ 2-photon transition is also negligible due to the $\pi$ transition symmetry. We thus set Raman coherence $\rho_{a b}=0$. For atoms subjected to a train of $j=1,...,N_{\rm total}$ shaped pulses, the state dynamics described by Eq.~(\ref{equMaster}) is therefore reduced to the $\rho_{a a}^{(j)}$, $\rho_{b b}^{(j)}$ population dynamics. Here $\rho_{a a}^{(j)}$, $\rho_{b b}^{(j)}$ are the initial population in the $|a\rangle$, $|b\rangle$ ground states right before the $j^{\rm th}$-pulsed excitation. For each single sub-pulses with duration $\tau$, we assume a flat spectrum $\propto I_0(\omega)$ to excite $|a\rangle-|e\rangle$, $|b\rangle-|e\rangle$ transition uniformly. We consider $\Theta_0=\sum_{i=1}^N |\int \kappa {\bf E}_{i,{\rm out}}(t) \cdot {\bf d}_{e g} e^{i\omega_{eg}t} d t/\hbar|$, with a simplified $d_{eg}=d_{e a}=d_{e b}$, and with $S(\omega)=|s(\omega)|^2/(|s(\omega)|^2)_{\rm max}$ as in the main text. We thus have the pulsed excitation probability  
\begin{equation}
    p_e^{(j)}=\frac{|\Theta_0|^2}{4}\big(\rho_{a a}^{(j)}S(\omega_{eg}+\pi\delta f_{{\rm HFS},g})+\rho_{b b}^{(j)} S(\omega_{eg}-\pi\delta f_{{\rm HFS},g})\big).\label{eqEXIT}
\end{equation}
Here in the qualitative model we have set $\Omega_a=\Omega_b$ without losing generality. With $T_{\rm rep}\Gamma_{a,b}\gg1$, the ground-state redistribution of atomic population after the spontaneous emission $\Delta \rho_{a a}^{(j+1)}=\rho_{a a}^{(j+1)}-\rho_{a a}^{(j)}$ is given by

\begin{equation}
    \begin{array}{l}
    \Delta \rho_{a a}^{(j+1)} =\frac{|\Theta_0|^2}{4} \big(
    -S(\omega_{eg}+\pi\delta f_{{\rm HFS},g})\rho_{aa}^{(j)}\frac{\Gamma_b}{\Gamma_a+\Gamma_b}+\\
    ~~~~~~~~~~~~~~~~~~~~~~~~~~~~~~~S(\omega_{eg}-\pi\delta f_{{\rm HFS},g})\rho_{bb}^{(j)}\frac{\Gamma_a}{\Gamma_a+\Gamma_b}\big).
    \end{array}
    \label{eqEXIT2}
\end{equation}

For a train of pulses with sufficiently large $N_{\rm total}$, we are interested in the ``steady state''  $\rho_{a a}^{(ss)}$, $\rho_{b b}^{(ss)}$ with vanishing population redistribution by Eq.~(\ref{eqEXIT2}). Accordingly, the excitation probability $p_e^{(ss)}$ is given by Eq.~(\ref{eqEXIT}), but with $\rho_{aa}^{(j)}$, $\rho_{bb}^{(j)}$ replaced by  $\rho_{a a}^{(ss)}$, $\rho_{b b}^{(ss)}$. We set $\Gamma_a=\Gamma_b$ in the simple model, and have 
\begin{equation}
p_e^{(ss)}= \frac{|\Theta_0|^2}{4} \tilde S(\omega_{eg},\pi\delta f_{{\rm HFS},g}),
\label{equgammaSS}
\end{equation}
with spectral response function which depends on the shaped pulse spectrum factor $S(\omega)$ through
\begin{equation}
\tilde S(\omega,\delta f)=\frac{2 S(\omega-\pi\delta f)S(\omega+\pi\delta f)}{S(\omega-\pi\delta f)+S(\omega+\pi\delta f)}.\label{EqspecResponse}
\end{equation}

Notice $\tilde S(\omega,\delta f)$ is reduced to $S(\omega)$ at the small $\delta f$ limit for Eq.~(\ref{equgammaSS}) to recover the 2-level result in the main text. To see why $\tilde S(\omega, \delta f)$ may suggest ``super-resolved'' frequency response, we consider shaped pulse with $S(\omega)=0$ for $\omega_{eg}+  \pi \delta f_{{\rm HFS},g}$, but with significant $S(\omega_{eg}-\pi \delta f_{{\rm HFS},g})$. That is, the shaped pulse can drive $|b\rangle-|e\rangle$ transition efficiently, but do not has the frequency component to drive the $|a\rangle-|e\rangle$ transition. In contrast to a naive expectation that $p_e$ would merely be reduced by a $(|a\rangle-|e\rangle)$-weighted factor, we have $\tilde S(\omega_{eg},\delta f_{{\rm HFS},g})$ completely vanishes, suggesting atoms do not response to the light excitation at all. This is because the ground state is repopulated to $\rho^{(ss)}_{aa}=1$ so as to be completely ``dark'' to the excitation. Similar situation appears for $S(\omega_{eg}-\pi \delta f_{{\rm HFS},g})=0$ with significant $S(\omega_{eg}+\pi \delta f_{{\rm HFS},g})$. We thus expect two zeros of $p_e^{(ss)}$ when a zero of $S(\omega)$ is scanned between $\omega_{eg}\pm \pi \delta f_{{\rm HFS},g}$, regardless of how wide the $S(\omega)$ distribution is! This population redistribution leads to apparent ``super-resolved'' feature as in Figs.~\ref{fig:TraplossFull}. The features demonstrate high resolution of the programmable shaped pulse at GHz-level in this work, with passive waveform stability over many hours (Though slow waveform drifts due to drifts of optical alignments are still expected, as suggested by the data and discussed in Sec.~\ref{SecWFStability}.).





\begin{thebibliography}{99}
\bibitem{notesFshift}
More precisely we have ${\bf E}_{\rm out}(t)=\kappa \sum_i^N A_i^2 e^{i \varphi_i } {\bf E}_{\rm in}(t+\tau_i)e^{-2\omega_i t}$, but the phase shift developed by the $e^{-2\omega_i t}$ factor is negligible for single picosecond pulses.
\end{thebibliography}
\end{document}